# Quantitative Geometric Market Structuralism (QGMS)

# A Framework for Detecting Structural Endpoints in Financial Market

**Author:** Amir Kavoosi

**Affiliation:** Independent Researcher

**Date:** November 2025

**Email:** Quantitative.price.action@gmail.com

**LinkedIn:** https://www.linkedin.com/in/amir-kavoosi-836a00378

**QGMS Research Series**

**Paper I: Theoretical Framework**

This paper is part of the Quantitative Geometric Market Structuralism (QGMS) Research Series, a multi-stage independent research program exploring geometric-quantitative approaches to market structure analysis.

# Abstract

This study introduces the Quantitative Geometric Market Structuralist (QGMS) framework — a hybrid analytical methodology integrating geometric pattern recognition with quantitative mathematical modeling to identify terminal zones of large-scale market movements. Unlike conventional econometric or signal-based models, the QGMS framework conceptualizes market dynamics as evolving geometric structures governed by self-organizing principles of price formation.

To preserve the proprietary nature of its internal mathematical architecture, the methodology employs a blind-testing validation process, wherein price, symbol, and temporal identifiers are concealed during analysis. This design ensures objective verification without revealing the underlying algorithmic core. The framework's predictive robustness has been empirically examined across multiple financial crises, including the 2008 Global Financial Collapse, the 2015 EUR/CHF SNB event, the 2016 Brexit referendum, and the 2020 COVID-19 market crash. In each case, the system consistently identified structural endpoints preceding major market reversals.

The findings suggest that geometric–quantitative market interpretation may offer a new class of predictive tools bridging the gap between mathematical formalism and empirical price behavior. By combining academic testability with intellectual property protection, the QGMS framework establishes a viable foundation for institutional evaluation and further research into non-linear structural forecasting models.

TABLE OF CONTENT

## Contents

# 1. Introduction:

Financial markets exhibit recurrent structural transformations that often precede large-scale price reversals. Traditional econometric and technical frameworks—while effective in capturing short-term correlations—frequently fail to detect deeper geometric symmetries governing market turning points. These symmetries, often embedded in the multi-scale architecture of price evolution, suggest that markets may self-organize according to underlying structural laws rather than stochastic randomness alone.

The 2008 Global Financial Collapse, the 2015 Swiss National Bank (SNB) event, the 2016 Brexit referendum, and the 2020 COVID-19 crash all exposed the limitations of conventional linear forecasting models. Despite abundant econometric data, few analytical systems successfully identified the structural exhaustion zones that preceded these reversals. This persistent predictive gap indicates a need for a new interpretative paradigm—one capable of decoding markets not as statistical aggregates but as evolving spatial constructs.

In response, the Quantitative Geometric Market Structuralist (QGMS) framework was developed to bridge this gap. QGMS integrates

principles of geometric morphology and quantitative validation into a unified model that treats price evolution as a dynamic geometric process. Instead of relying solely on lagging statistical relationships, it emphasizes the intrinsic geometry of market movements, focusing on proportional, angular, and temporal harmonics that signal structural convergence or divergence.

To ensure methodological credibility while preserving proprietary integrity, the framework employs a blind-testing validation protocol, in which all price, time, and symbol identifiers are concealed prior to structural analysis. This enables objective assessment of predictive performance without disclosing the internal computational logic.

The present study outlines the conceptual foundations of QGMS, describes its validation process, and presents empirical evidence from major historical crises to demonstrate the framework's structural predictive capability. The findings aim to contribute to the broader discourse on non-linear market dynamics, offering a pathway toward the next generation of quantitative-structural forecasting models.

At its core, the QGMS framework assumes that price movements can be decomposed into geometric substructures whose internal relationships obey deterministic proportional and angular constraints.

Each structural segment carries a quantifiable geometric signature, and these signatures must remain hierarchically coherent across multiple scales of market evolution. Structural turning points emerge not from stochastic volatility clusters but from the exhaustion of these inter-segment geometric relationships — a state referred to in this framework as structural saturation.

## 2. Theoretical Foundation:

The QGMS framework is built upon two complementary theoretical pillars that describe how market structures evolve and where they terminate.

### 2. 1 Geometric Structuralism:

Price movements are interpreted as sequences of geometric substructures rather than random fluctuations. Each substructure carries a distinct geometric signature—defined by its proportional, angular, and directional characteristics—which must remain internally coherent with adjacent structures.

This framework assumes that market trends develop geometric tension as these structures interact. When the tension reaches its saturation limit, the trend becomes structurally incapable of further continuation, producing a deterministic reversal point.

## 2. 2 Quantitative Structural Convergence:

To formalize these geometric relationships, each structural segment is transformed into a normalized quantitative coefficient. These coefficients must satisfy multi-scale consistency constraints: small-scale structures must remain compatible with the geometric signature of larger ones.

Although the specific encoding mechanism is proprietary, the operational principle is testable: when the geometric coefficients across scales converge toward their allowable boundary, the system identifies a structural termination zone.

Unlike classical econometric models, which rely on probabilistic inference, this framework asserts that the governing law of trend completion emerges from internal geometric exhaustion rather than external randomness.

# 3. Methodology:

## 3. 1 Framework Overview

The Quantitative Geometric Market Structuralist (QGMS) framework operates on the premise that market price trajectories evolve through discrete structural phases governed by internal geometric proportions rather than continuous stochastic drift. Each phase exhibits a measurable internal symmetry—expressed through spatial ratios, angular harmonics, and temporal intervals—that collectively define a market's "structural signature."

At its core, the QGMS model identifies terminal zones, regions where the internal symmetry of the market's geometric configuration approaches exhaustion. These terminal zones represent areas of potential trend culmination, often preceding significant reversals in price direction. The system's interpretive logic relies on geometric coherence rather than directional prediction—focusing on where the structure completes rather than which way the market will move.

Unlike traditional econometric models that derive conclusions from regression or correlation, QGMS interprets market data as a dynamic field of geometric interactions. The analytical process unfolds in three conceptual stages:

**1. Structural Decomposition** – Deconstruction of the price evolution into discrete geometric components across multiple temporal scales.

**2. Geometric Convergence Analysis** – Identification of proportional and angular relationships suggesting the exhaustion or alignment of structural patterns.

**3. Terminal Zone Validation** – Cross-comparison of geometric endpoints against blind datasets to assess predictive validity without revealing underlying formulaic parameters.

To protect the proprietary nature of its algorithmic logic, QGMS does not disclose explicit mathematical functions or variable transformations. Instead, its credibility is established through blind-testing protocols, where all market identifiers—such as asset symbol, price scale, and timeframe—are concealed during analysis. The structural predictions are then compared to real post-event outcomes, allowing objective verification of accuracy while preserving the internal mechanism's confidentiality.

This dual-layer design—combining conceptual transparency with technical protection—ensures that QGMS remains both scientifically testable and intellectually secure. It enables academic evaluation based on reproducible validation procedures rather than direct algorithmic exposure, aligning the framework with ethical standards of empirical research while safeguarding its intellectual property.

## 3. 2 Blind Test Procedure

To validate the QGMS framework while preserving the proprietary nature of its internal mathematical architecture, a blind-testing protocol was designed. The procedure consists of the following stages:

**Technical Specifications:**
**1. Chart Preparation:** Each chart is locked from a specified start date, with complete removal of all price values, symbols, and date identifiers.

Only the candlestick structure and corresponding time frame are retained.

**2. Forward Analysis:** Charts are presented in a manual, candle-by-candle forward manner to the analyst, without revealing any historical outcomes.

**Stage 1 – Blind Analysis:**

The analyst performs the structural and geometric assessment of the market solely based on candlestick formations and time frame information, with no knowledge of the underlying asset, or specific dates.

**Stage 2 – Quantitative Evaluation:**

The generated signals are executed on the evaluator's platform (e.g., MetaTrader Strategy Tester) by an independent testing team. Performance metrics, including drawdown and risk-to-reward ratio (R/R), are computed and reported. Professional evaluation reports are provided immediately upon completion of the blind test, ensuring objective verification of the methodology without disclosure of proprietary computations.

**This two-stage blind testing protocol ensures that:**

1. All predictions are generated and recorded in a controlled sequence prior to outcome evaluation
    1. Performance metrics are calculated independently
    2. The core mathematical architecture remains fully protected
    3. The validation process meets academic rigor standards.

# 4. Structural Encoding Formalization

The QGMS framework defines a deterministic transformation that converts raw price data into a sequence of structural coefficients representing the geometric organization of market movement. The procedure consists of two main stages:

**1. Segmentation**

Price series → S = [S1, S2, …, Sn]

The price series is decomposed into a sequence of geometric segments, each representing a coherent structural phase such as impulsive expansion, contraction, or corrective consolidation.

Segmentation isolates the internal geometric roles within the price pattern and prepares each unit for structural encoding.

## 2. Encoding

Segments → C = [c1, c2, …, cn]

Each segment Si is mapped into a structural coefficient through the encoding operator:

$$c_i = \Phi(S_i)$$

The operator Φ is proprietary, but its functional purpose is clear: it extracts a normalized, scale-invariant geometric signature from each structural segment.
The internal mathematical formulation of Φ remains undisclosed for intellectual property protection.

**Hierarchical Admissibility Constraint**

If a segment S_child is embedded within a larger segment S_parent, its coefficient must lie within the geometrically admissible coefficient space defined by the parent structure:

$$c\_child \in A(c\_parent)$$

Here, A(c_parent) does not require numerical similarity or identical proportional coefficients.
Instead, it represents a set of permissible geometric behaviors that preserve the structural role of the parent segment (for example, impulsive vs. corrective structure).

This ensures multi-scale compatibility without enforcing equality.

**Framework Properties**

1. Determinism

Given identical input segments, the encoding operator Φ always produces identical coefficients.
No stochastic or random component is involved.

## 2. Scale Robustness

If a structural segment is uniformly expanded or compressed by a scalar factor α, its encoded coefficient remains approximately invariant:

$$\Phi(\alpha \cdot S\_k) \approx \Phi(S\_k)$$

This reflects that QGMS encodes geometric form, not absolute price magnitude.

## 3. Hierarchical Consistency

Coefficients produced at different structural scales must satisfy admissibility relationships.

Local (child) structures cannot produce coefficients that contradict the geometric constraints imposed by parent structures.

## 4. 1 Definitions & Axiom

We define a structural segment as a contiguous geometric phase of price evolution characterized by consistent directional behavior and internally coherent sub-oscillation patterns.

> Each segment is transformed into a normalized structural coefficient through an encoding operator Φ, which maps geometric magnitude into a scale-invariant representation.

> The QGMS framework is governed by three axioms:

**1. Axiom A1 — Determinism**
For any structural segment, the encoding operator Φ produces a unique coefficient.
Identical inputs always yield identical outputs.

## 2. Axiom A2 — Scale Invariance

For any scalar factor α, uniformly scaling a geometric segment does not change its encoded coefficient.

In symbolic form:

$$\Phi(\alpha \times S) = \Phi(S)$$

## 3. Axiom A3 — Hierarchical Consistency

If a segment is contained within a larger segment, its coefficient must lie within the geometrically admissible coefficient space defined by the parent:

$$c_{child} \in A(c_{parent})$$

Here, does not impose numerical equality or proportional replication.

Instead, it defines a range of permissible geometric behaviors that preserve the structural role of the parent segment (e.g., impulsive, corrective, consolidative).

A child structure may possess its own multi-phase coefficient pattern provided its behavior remains compatible with the overarching geometric logic of the parent.

## 4. 2 Lemma Sketches: (Theoretical Consequences)

**Lemma 1** — Preservation of Structural Ordering

Under the Scale Invariance axiom (A2), the relative ordering of coefficients across sibling segments remains unchanged under uniform scaling of price.

**Proof Sketch:**

Because Φ operates on normalized geometry, the scaling factor α cancels during normalization.
Thus, proportional relationships among segments remain invariant, preserving the ordering, etc.

**Lemma 2 — Multi-Scale Structural Coherence**

Statement:

Under the Hierarchical Admissibility Constraint, if the parent coefficient approaches the boundary of its admissible region , then at least one child coefficient must simultaneously approach the boundary of its own admissible region.

**Proof Sketch:**

Because

$c_{child} \in A(c_{parent})$,

As the parent structure approaches geometric saturation, the permissible geometric configurations for its substructures shrink accordingly.

Thus, structural saturation propagates downward across scales, causing the entire multi-scale structure to collapse coherently at the terminal point.

# 5. Empirical Validation Overview:

To assess the predictive reliability of the QGMS framework, a series of retrospective blind tests were conducted across multiple macro-financial turning points, including the 2008 Global Financial Collapse, the 2015 EUR/CHF SNB event, the 2016 Brexit referendum, and the 2020 COVID-19 market crash.

In each case, the model operated under strict blind conditions — with asset names, temporal context, and event identifiers concealed — ensuring that results reflected pure structural interpretation, free from narrative bias.

Detailed results and analyses for each case are presented in companion papers within the QGMS Research Series, where each study explores a distinct structural configuration and market environment. Collectively, these validations demonstrate the framework's consistent capacity to identify terminal zones preceding major reversals across diverse asset classes and conditions.

## 5.1 Case Study: Structural Non-Convergence in USD/CHF

**Market Context & Setup:**

Price approached a significant technical confluence at the 0.84400 level, intersecting both a historical order block and the lower boundary of a descending trend channel. Despite this classical technical setup, the market exhibited a notably weak reaction, producing only minimal bullish momentum before continuing its downward trajectory.

**Structural Analysis Using QGMS Framework:**

The QGMS framework rejected this region as a high-probability reversal zone due to two critical structural incompatibilities:

**1. Geometric Disparity Between Legs AB and CD**
  · Segment AB displayed a compressed, rhythmically consistent decline with balanced internal oscillations
  · Segment CD exhibited a sharp, high-momentum descent characterized by expanding volatility and reduced structural coherence
  · The two legs failed to demonstrate the geometric homology required for valid trend termination.

**2. Quantitative Coefficient Misalignment**

· The proprietary encoding operator Φ produced coefficient values for AB and CD that occupied distinct regions within the structural coefficient space

· This mathematical discrepancy indicated fundamentally different structural signatures between the two segments

· Hierarchical validation checks confirmed the absence of multi-scale pattern alignment

**Framework Decision & Rationale:**

QGMS classified this structure as "geometrically incomplete" and generated no entry signal. The framework identified the apparent technical confluence as structurally invalid due to the underlying geometric contradictions.

**Ex-Post Validation:**

Subsequent price action confirmed the framework's assessment:

· Price briefly consolidated at 0.84400 before resuming its decline.
· The minimal reaction validated the absence of genuine structural exhaustion.

· The decline extended approximately 150 pips following the failed confluence.

**Comparative Advantage**

This case demonstrates QGMS's ability to filter false signals that would typically trigger entries in conventional technical analysis. By prioritizing structural completeness over superficial pattern recognition, the framework avoided a potentially costly misinterpretation of market structure.

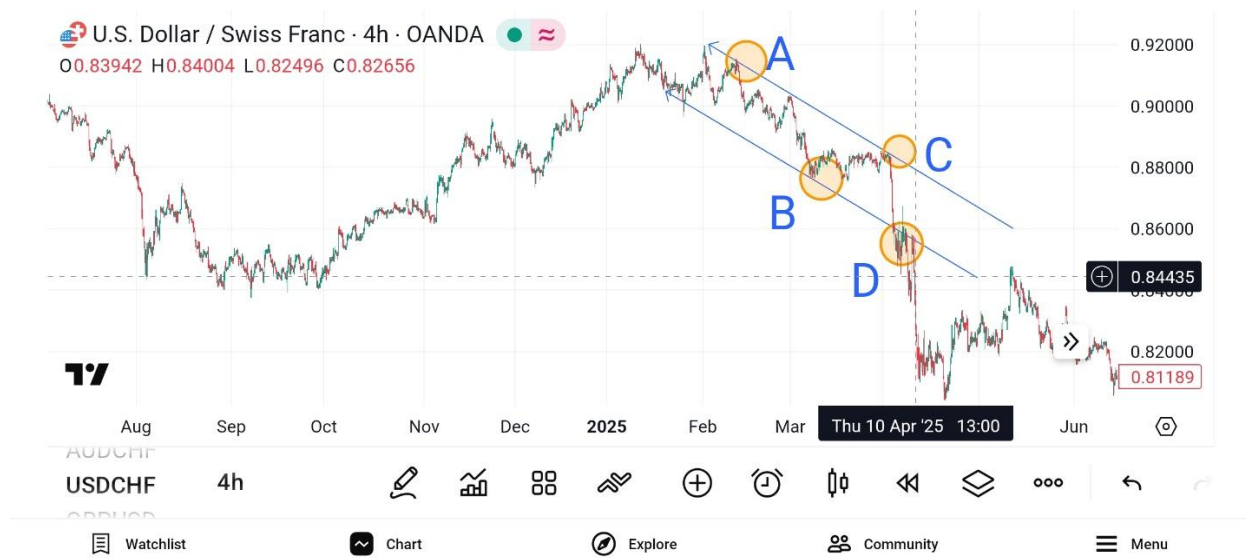

## 5.2 Case Study2: Geometric Identification of Trend Exhaustion in GB10Y

**Beyond Statistics: How Geometric Mathematics Identifies the End of a Trend (GB10Y Case Study)**

Conventional quantitative models are brilliant at measuring momentum and volatility. But they consistently fail to answer the most critical question: Where does the trend end?

Look at this 10-year UK Gilt (GB10Y) chart. Could any statistical model have pinpointed the precise reversal at point C before it happened? The answer is no. They are reactive by nature.

My work in Geometric Market Structuralism solves this. It's a new paradigm in quantitative analysis that moves beyond price data to decode the market's underlying mathematical structure.

Here's the framework in action on this chart:

**1. Deconstruct the Trend:** The macro structure (A to C) is broken down into two core sub-structures: AB and BC.

**2. Identify Dynamic Relationships:** A coherent mathematical relationship, governed by dynamic geometric coefficients, is calculated between these three structures (AB, BC, and the superstructure AC).

**3. Pinpoint the Terminal Point:** The BC sub-structure is identified as the terminating structure. As price approaches point C, the model detects

Structural Saturation—the point where the mathematical relationship between all structures reaches its conclusion.

The result? The endpoint at C is not a probabilistic forecast; it is a pre-defined mathematical conclusion derived from the market's own architecture.

This is the same methodology that defined the endpoint of the 2015 SNB catastrophe in real-time. It is not pattern recognition; it is structural causality.

For portfolio managers and risk officers: Are your current models equipped to see the end, or just follow the path?

NOTE: "A critical distinction: This is not Elliot Wave theory. Where Elliot Wave relies on subjective wave counts and fixed Fibonacci ratios, this framework is based on objective, dynamic geometric coefficients that calculate a definitive endpoint through structural saturation. It's mathematics, not psychology.

"The proprietary dynamic coefficients and the exact mathematical definition of structural saturation are, for obvious reasons, the protected intellectual property of this methodology.

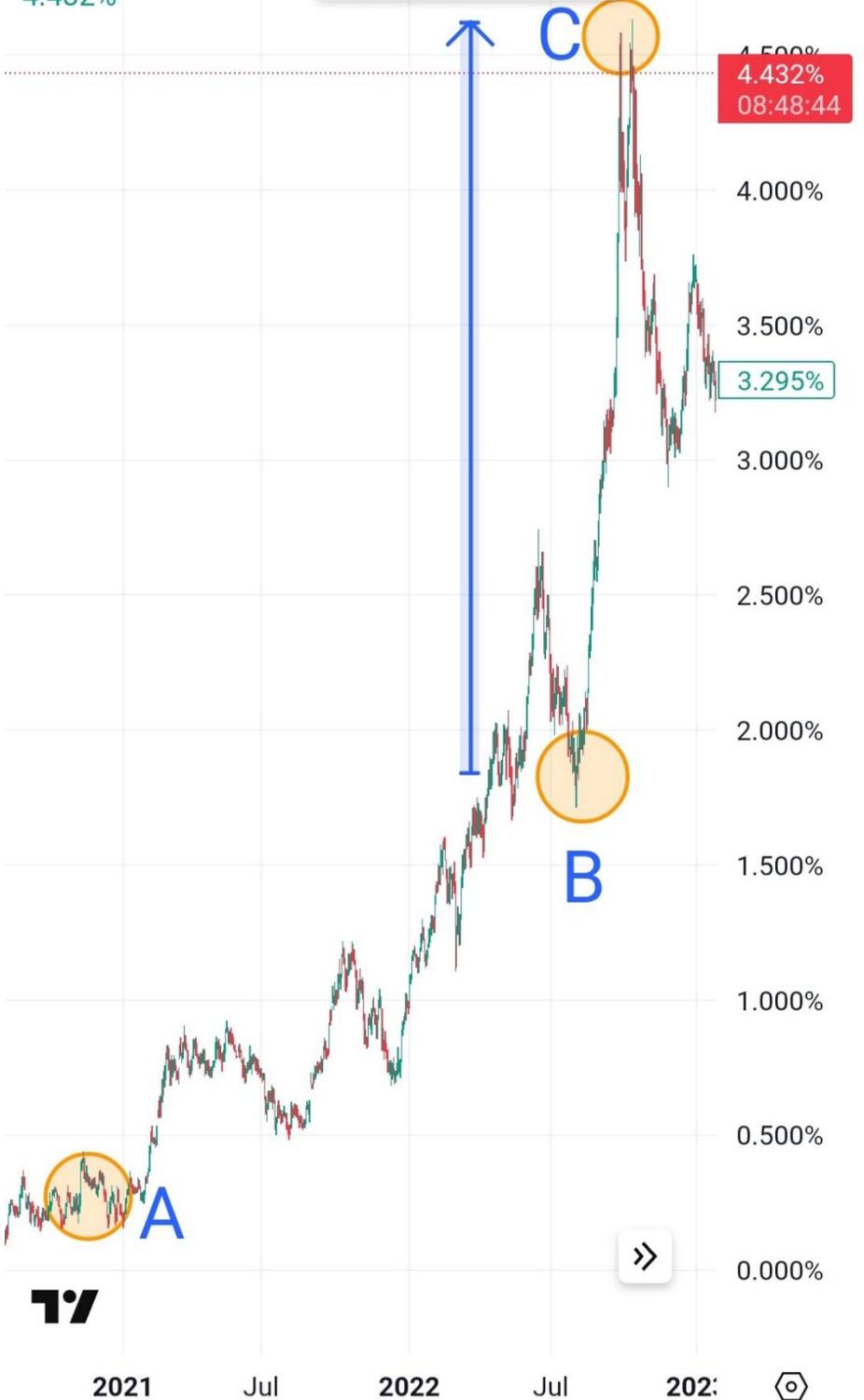

## 5. 3 Case Study: Structural Resolution of the 2015 SNB Crisis:

The Limitations of Conventional Risk Models

The 2015 Swiss National Bank event demonstrated the critical failure point of traditional risk management systems. As EUR/CHF plummeted approximately 2300 pips, standard volatility-based models proved incapable of identifying the structural termination zone.

**QGMS Framework Performance**

In contrast, the geometric structural analysis identified the precise exhaustion point in real time by decoding multi-scale structural patterns that had been evolving since October 2010.

**Methodological Validation**

This case represents one instance of a comprehensive validation process spanning

25 years of market data, including the 2008 Global Financial Crisis, Brexit

flash crash, and COVID-19 market collapse.

**Structural Opportunity Identification**

The accurately identified termination point revealed a reversal potential exceeding 1000 pips, demonstrating the framework's capacity to convert structural analysis into measurable forecasting advantage.

**Implications for Risk Management**

This case study suggests that geometric structural analysis may offer institutions a complementary approach to conventional risk modeling,

particularly during periods of extreme market discontinuity.

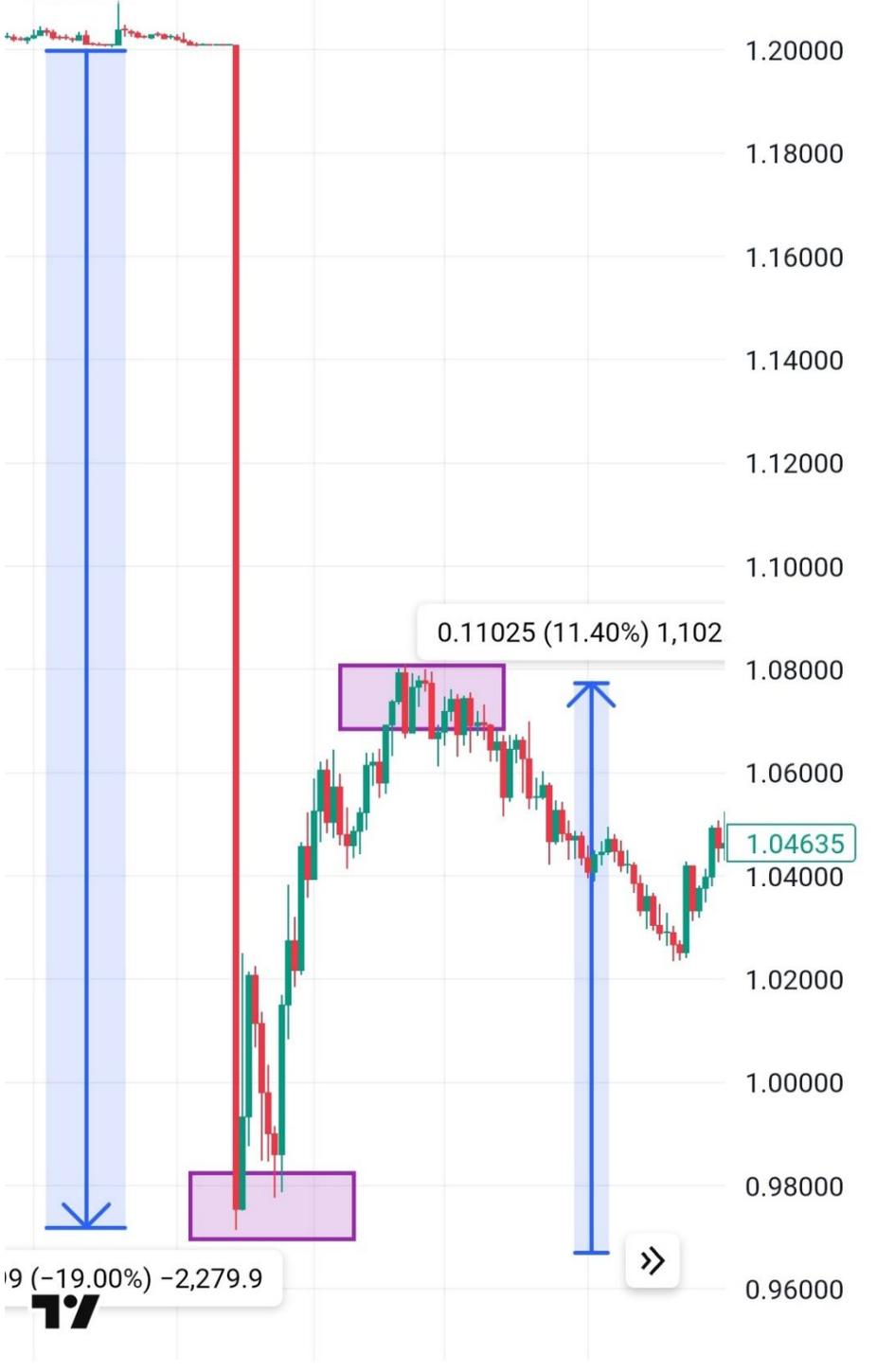

# 6. Discussion and Practical Validation

The practical validation of the QGMS framework rests on its ability to be empirically tested without disclosure of its proprietary mathematical core. Unlike conventional quantitative systems that require source-code or algorithmic access, the QGMS methodology has been designed to operate within a blind analytical protocol — a controlled environment that allows independent verification of performance through observation rather than replication.

## 6. 1 Validation Philosophy

Conventional academic replication demands full algorithmic transparency. However, in proprietary research fields such as quantitative finance, intellectual property protection must coexist with scientific accountability. The QGMS approach reconciles these objectives through observational reproducibility: third parties can evaluate outputs and logic coherence even without reconstructing internal equations.

This dual integrity — confidentiality with verifiability — positions QGMS as a bridge between academic methodology and real-world proprietary systems.

## 6. 2 Blind-Testing Framework

The proposed validation protocol is structured into two independent stages:

**Stage I** — Structural Analysis (Independent Analyst)

A locked chart sequence is provided in which all identifying information — including symbol, date, and price scale — is removed.

The analyst applies the QGMS method purely on the geometric and temporal structure, identifying potential terminal zones.

**Stage II** — Quantitative Verification (Independent Institution)

The signals from Stage I are executed within an institutional testing environment (e.g., MetaTrader, MATLAB, or a proprietary backtester).

Metrics such as drawdown, success ratio, and structural accuracy are calculated.

This two-layer separation ensures that validation remains unbiased, mathematically measurable, and reproducible in principle.

## 6. 3 Transparency and Integrity

All analytical outputs are timestamped and documented before decoding to prevent hindsight bias.
The blind framework allows evaluators to:

1. Verify that predictions were made before event revelation.
2. Assess structural accuracy quantitatively.
3. Maintain full confidentiality over the internal logic of the model.

Thus, QGMS promotes verifiable transparency without disclosure — an essential equilibrium for independent research operating at the intersection of mathematics and market dynamics.

## 6. 4 Institutional Application

This validation architecture allows institutional or academic collaborators to test the framework under real data conditions without any compromise of intellectual ownership.
Potential areas for future collaboration include:

1. Comparative analysis with econometric volatility models,
2. Integration into risk management algorithms,
3. Academic joint studies on geometric predictability and self-organizing market structures.

By emphasizing methodological clarity and testability, the QGMS framework invites critical examination — not as a closed black-box, but as a mathematically reasoned model that can be evaluated empirically and ethically.

# 7. Conclusion and Future Work

This study introduced the Quantitative Geometric Market Structuralist (QGMS) framework — a novel analytical paradigm that interprets market behavior through the intersection of geometry and adaptive mathematics.

By reconceptualizing price movement as a self-organizing geometric system rather than a stochastic process, the QGMS framework establishes a foundation for a new class of predictive tools that are both quantitative and structural in nature.

The blind-testing validation design ensures that the framework remains scientifically testable while protecting the proprietary mathematical architecture that underpins its logic.
Through this balance, the study contributes a replicable verification path for independent researchers and institutions seeking to evaluate complex systems without direct access to source algorithms.

From a broader perspective, the results demonstrate that financial market reversals and terminal zones may not be purely random phenomena but instead follow measurable geometric and mathematical relationships that evolve dynamically across time.
This insight opens new directions for academic investigation into nonlinear structural predictability, potentially bridging the divide between theoretical finance and practical algorithmic application.

## 7. 1 Future Research Directions:

**1. Comparative Analysis with Econometric Models –**

Future studies could integrate the QGMS framework with volatility-based econometric approaches (e.g., GARCH, SV models) to examine potential synergies between structural geometry and statistical variance modeling.

**2. Expansion Across Asset Classes –**

While initial validation has focused on major currency pairs and commodities, the same principles could be extended to equity indices, bonds, and emerging cryptocurrency markets to assess cross-domain structural consistency.

**3. Mathematical Formalization of Geometric Ratios –**

Continued research may focus on translating the geometric structures into partially public mathematical expressions, offering a bridge between proprietary modeling and academic formalism.

## 7. 2 Closing Remark

The QGMS framework is not presented as an ultimate or closed system, but as an evolving research foundation.

Its purpose is to inspire collaboration between academia and proprietary research — to prove that rigorous mathematical reasoning can co-exist with intellectual property protection.

By offering a transparent path for blind empirical validation, this study invites constructive engagement from scholars, institutions, and market researchers who share the goal of uncovering the deeper structural logic of financial dynamics.

**Further case-specific analyses expanding on this theoretical foundation will be presented in subsequent companion papers within the QGMS Research Series.**

## References

This study is entirely based on the author's original research and proprietary analytical framework. No external sources were used.